\documentclass[final,conference, twoside, 10pt, letterpaper]{IEEEtran}
\usepackage[hyperref]{Bamble}
\usepackage{Bmacros}
\usepackage[noadjust]{cite}

\newcommand{\twobibs}[2]{#2} 

\begin{document}
\title{List Decoding of Universal Polar Codes}
\date{}
\author{\IEEEauthorblockN{Boaz~Shuval, Ido~Tal\\
Department of Electrical Engineering,\\
Technion, Haifa 32000, Israel.\\
Email: \{\texttt{bshuval@}, \texttt{idotal@ee.}\}\texttt{technion.ac.il}}}

\maketitle

\begin{abstract}
    A list decoding scheme for universal polar codes is presented. Our scheme applies to the universal polar codes first introduced by \sasoglu and Wang, and generalized to processes with memory by the authors. These codes are based on the concatenation of different polar transforms: a sequence of ``slow'' transforms and \arikan's original ``fast'' transform. List decoding of polar codes has been previously presented in the context of the fast transform. However, the slow transform is markedly different and requires new techniques and data structures. We show that list decoding is possible with space complexity $O(\listSize \cdot N)$ and time complexity $O(\listSize \cdot N \log N)$, where $N$ is the overall blocklength and $\listSize$ is the list size.
\end{abstract}

\section{Introduction}
\looseness=-1
Polar codes, introduced by \arikan~\cite{Arikan_2009}, are a rich a family of codes. 
They have been extended to many settings,
e.g.,~\cite{sasoglu_nonbinary,Korada_source,arikan_2010_source,Hof_wiretap,Korada_2010_kernel,Mahdavifar_wiretap,Honda_Yamamoto_2013,Hassani_universal,sasoglu_2016_universal,Tal_Memory_2019,Shuval_Tal_Memory_2019,Goldin_2019,Tal_Pfister_Deletion}.
Specifically, in~\cite{sasoglu_2016_universal}, a universal polar coding construction was presented.
It applies to a setting where the channel is unknown to the encoder, but the decoder has full
channel knowledge, obtained, e.g., by channel estimation (see also~\cite{Hassani_universal} for another approach). 
In~\cite{Shuval_Tal_Universal}, the construction of~\cite{sasoglu_2016_universal} was generalized to apply to a large class of channels and sources with memory.
List decoding is a technique to improve upon the successive-cancellation (SC) decoding error of polar codes. 
Its implementation in~\cite{Tal_2015} applies to \arikan's seminal polar codes~\cite{Arikan_2009} 
(see also~\cite{Dumer_list}), but does not trivially extend to the universal construction. 
In this paper, we present such an extension. 

\looseness=-1
The universal polar codes of~\cite{sasoglu_2016_universal,Shuval_Tal_Universal} are based on concatenating recursive transforms of two types.
One of which, called the \emph{fast} transform, is \arikan's transform~\cite{Arikan_2009}; the
other, called the \emph{slow} transform, is different. 
A key step in this concatenation is joining multiple copies of a slow transform such that their outputs are fed into multiple fast transforms. 
Both transforms above can be described using \emph{layers}.  Specifically, the layer keeps track of the recursion depth. 

\looseness=-1
Let $x_0^{N-1}$ denote the transform input and $u_0^{N-1}$ its output. 
Consider first the fast transform~\cite{Arikan_2009}. 
Here, decoding is performed via successive-cancellation, in which symbols are decoded successively.
Namely, we decode symbol $u_i$ after having decoded the previous symbols $u_0^{i-1}$ and under the assumption that our previous decoding decisions $\hat{u}_0^{i-1}$ are correct. 
The slow transform can also be decoded using successive-cancellation. 
A crucial property of SC decoding of the fast transform is that in decoding symbol $u_i$, typically only a small number of layers are involved. 
In stark contrast, as we will later see, in SC decoding of the slow transform, typically \emph{all} layers are involved. 

Successive-Cancellation List (SCL) decoding is a generalization of SC decoding, in which multiple ``decoding paths'' are considered in parallel. 
That is, instead of making a hard decision on the value of $\hat{u}_i$ at stage $i$, we allow for multiple hypotheses. 
We keep the list of hypotheses manageable by pruning it and keeping the $\listSize$ most-likely paths. 
In a fast transform of blocklength $N$ this can be accomplished in space $O(\listSize\cdot N)$ and time $O(\listSize \cdot N \log N)$. 
The crucial property highlighted in the previous paragraph is what enables this for the fast transform. 
The crucial property does not hold for the slow transform.
Thus, different techniques are needed to accomplish SCL decoding with the same space and time complexity for the universal concatenated transform. We indeed accomplish this, by exploiting a certain cyclic property of the slow transform. We leverage this property through a dedicated data-structure, termed the \emph{cyclic exponential array}.

\looseness=-1
Due to length constraints, we focus on the slow transform and only outline the entire construction.
Proofs and numerical results are omitted. 
A full paper with all the details is in preparation.

\section{The Slow Transform}\label{sec_slow}
The slow transform introduced in~\cite{sasoglu_2016_universal} was streamlined and generalized to settings with memory in~\cite{Shuval_Tal_Universal}. 
We now describe the slow transform of~\cite{Shuval_Tal_Universal}.
Unlike~\cite{Shuval_Tal_Universal}, We use zero-based indexing, as it is more amenable to implementation. 

\looseness=-1
The slow transform is one-to-one and onto. 
It transforms a vector of $N$ bits into another vector of $N$ bits. 
The transform is recursively defined. 
The initial step is an identity mapping of length $N_0 = 2L_0 + M_0$. 
Parameters $L_0$ and $M_0$ are selected according to the memory properties of the setting and the desired rate, see~\cite{Shuval_Tal_Universal}.
Every step of the recursion doubles the transform length. After $n$ steps it transforms vectors of length $N=2^n N_0$.

\looseness=-1
Borrowing terminology from~\cite{Tal_2015}, we describe a slow transform of recursion depth $n$
using \emph{layers},  \emph{phases}, and \emph{branches} as follows.
A slow transform of size $N = 2^n N_0$ --- i.e., of recursion depth $n$ --- has $n+1$ layers, from $0$ to $n$.  
Layer $0$ is associated with the transform input $x_0^{N-1}$; layer $n$ is associated with the transform output $u_0^{N-1}$. 
Each layer of the transform is divided into branches, each comprising a contiguous set of indices called phases. 
The number of branches in layer $\lambda$ is $2^{n-\lambda}$; each branch comprises $2^{\lambda}
N_0$ phases. 
The mapping between index $i$ in layer $\lambda$ and its  phase $\varphi$ and branch $\beta$ is
\[ i = \pbl{\varphi}{\beta}{\lambda} \triangleq \varphi + \beta \cdot 2^{\lambda}N_0, \quad 0\leq
\varphi < 2^{\lambda} N_0, \quad 0\leq\beta < 2^{n-\lambda}  .\] 
When the layer $\lambda$ is obvious from the context, we omit it and simply write $i=\pb{\varphi}{\beta}$.

We divide the phases within a branch $\beta$ of any layer $\lambda$ to several sets. 
As in~\cite{Shuval_Tal_Universal}, we assume that $M_0$ is even, and define
\begin{equation} 
    L_{\lambda} = 2^{\lambda}(L_0 +1) -1, \quad M_{\lambda} = 2^{\lambda} (M_0 - 2) + 2.
 \label{eq_Ln}
\end{equation} 
In other words, $L_{\lambda} = 2L_{\lambda-1}+1$ and $M_{\lambda} = 2(M_{\lambda-1} -1)$.  
Observe that $2L_{\lambda}+M_{\lambda}=2^{\lambda}N_0 \triangleq N_{\lambda}$, the number of phases in a branch of layer
$\lambda$. 
The lateral set, $\latset{\lambda}$, consists of $2L_\lambda$ phases. 
The medial set, $\medset{\lambda}$, consists of the remaining $M_\lambda$ phases. 
These sets are further subdivided. 
The mapping of phases to sets in any branch of layer $\lambda$ is given by: 
\begin{IEEEeqnarray}{rCl} \IEEEyesnumber \label{eq_defs of lat-med sets}
    \latsettop{\lambda} &\triangleq& \{\varphi \ | \ \hphantom{L_{\lambda}+M_{}} 0 \leq \varphi \leq L_{\lambda}-1\},  \IEEEyessubnumber \label{eq_def of lat1}\\
    \medsetMinus{\lambda}   &\triangleq& \{\varphi \ | \ \varphi = L_{\lambda} + 2k,\hphantom{{}+1} \ 0 \leq
    k < M_{\lambda}/2\},  \IEEEeqnarraynumspace \IEEEyessubnumber \label{eq_def of medA} \\    
    \medsetPlus{\lambda}   &\triangleq& \{\varphi \ | \ \varphi = L_{\lambda} + 2k+1, \ 0 \leq k < M_{\lambda}/2\},  \IEEEyessubnumber \label{eq_def of medB}\\
    \latsetbot{\lambda} &\triangleq& \{\varphi \ | \  L_{\lambda} + M_{\lambda} \leq \varphi \leq N_{\lambda}-1\},  \IEEEyessubnumber \label{eq_def of lat2} \\
    \latset{\lambda}    &\triangleq& \latsettop{{\lambda}} \cup \latsetbot{{\lambda}},  \IEEEyessubnumber \label{eq_def of latset} \\
    \medset{\lambda}    &\triangleq& \medsetMinus{n} \cup \medsetPlus{n}.  \IEEEyessubnumber \label{eq_def of medset}
\end{IEEEeqnarray}                              
Observe that the first $L_{\lambda}$ phases of a branch are lateral, the next $M_{\lambda}$ phases
are medial and alternate between $\medsetMinus{\lambda}$ and $\medsetPlus{\lambda}$, and the final
$L_{\lambda}$ indices are again lateral. 

Let $x_0^{N-1}$ be the input and $u_0^{N-1}$ the output of a slow transform of recursion depth $n$. 
Denote index $i = \pbl{\varphi}{\beta}{\lambda}$ of the vector corresponding to layer $\lambda$ by
$u^{(\lambda)}_i = u^{(\lambda)}_{\pb{\varphi}{\beta}}$. 
In particular, $u^{(0)}_i = x_i$ and $u^{(n)}_i =u_i$. 
We also denote 
\vspace{-0.1cm}
\[ \psi = \left\lfloor\frac{\varphi}{2}\right\rfloor, \quad \psi' = \left\lfloor\frac{\varphi-1}{2}\right\rfloor.\]  
Then, the slow transform recursion for $\lambda \geq 1$ is given by 
\begin{IEEEeqnarray}{rCl} 
    \varphi \!\in\! \latset{\lambda} \!\Rightarrow\!
    u^{(\lambda)}_{\pb{\varphi}{\beta}} &\!=\!& \begin{cases}
        u^{(\lambda-1)}_{\pb{\psi}{2\beta}}\,, & \varphi \text{ even}, \\[0.3cm]
        u^{(\lambda-1)}_{\pb{\psi}{2\beta+1}}\,, & \varphi \text{ odd}, 
    \end{cases}  \label{eq_Fj for lateral indices_pbl} \\[0.2cm]
        \varphi \!\in\! \medset{\lambda}  \!\Rightarrow\!
        u^{(\lambda)}_{\pb{\varphi}{\beta}}  &\! =\! & \begin{cases} 
    u^{(\lambda-1)}_{\pb{\psi'+1}{2\beta}} + u^{(\lambda-1)}_{\pb{\psi'}{2\beta+1}}\,,\;\;\, \varphi
    \text{ odd,} &{} \\[0.45cm] 
    u^{(\lambda-1)}_{\pb{\psi'}{2\beta+1}}\,,\;\;\, 
    \substack{\mathmakebox[\widthof{$\displaystyle \psi' \!\in\!  \medsetMinus{\lambda-1},$}][l]{\displaystyle \varphi \text{ even, }} \\[0.1cm]
    \displaystyle \psi' \!\in\! \medsetMinus{\lambda-1},} & {} \\[0.45cm] 
    u^{(\lambda-1)}_{\pb{\psi'+1}{2\beta}}\,,\;\;\, 
    \substack{\mathmakebox[\widthof{$\displaystyle \psi' \!\in\!  \medsetPlus{\lambda-1}.$}][l]{\displaystyle \varphi \text{ even, }} \\[0.1cm]
    \displaystyle \psi' \!\in\! \medsetPlus{\lambda-1}. }
\end{cases}         \label{eq_Fj for medial indices_pbl}
\end{IEEEeqnarray}
Observe from \eqref{eq_Ln}, \eqref{eq_def of medA}, and \eqref{eq_def of medB}
that since $\lambda \geq 1$, $\varphi \in \medset{\lambda}$ is even if and only if $\varphi \in
\medsetPlus{\lambda}$. When  $\lambda \geq 2$, we have
\begin{align*} 
    \psi' \in \medsetMinus{\lambda-1} &\Longleftrightarrow \psi' \text{ is odd}, \\ 
    \psi' \in \medsetPlus{\lambda-1} &\Longleftrightarrow \psi' \text{ is even}. 
\end{align*}

By~\eqref{eq_Fj for lateral indices_pbl} and~\eqref{eq_Fj for medial indices_pbl}, branch $\beta$ of
layer $\lambda$ is formed from branches $2\beta$ and $2\beta+1$ of layer $\lambda-1$.
From~\eqref{eq_Ln} and \eqref{eq_defs of lat-med sets}, 
\emph{all} lateral phases of branches $2\beta$, $2\beta+1$ of layer $\lambda-1$ are 
transformed into lateral phases of branch $\beta$ of layer $\lambda$.  
Additionally, medial phases $\pbl{L_{\lambda-1}}{2\beta}{\lambda-1}$ and
$\pbl{L_{\lambda-1}+M_{\lambda-1}-1}{2\beta+1}{\lambda-1}$ become lateral phases of layer $\lambda$. 

The operation in~\eqref{eq_Fj for medial indices_pbl} consists of minus and plus transforms:
a minus transform for odd medial $\varphi$, and a plus transform for even
medial $\varphi$. 
Equation~\eqref{eq_Fj for medial indices_pbl} reveals a cardinal difference between the slow transform and \arikan's fast transform. 
In the fast case, the minus transform operates on the \emph{same} phase of two consecutive branches. 
In the slow case, the minus transform operates on \emph{consecutive} phases of two consecutive branches. 

\section{Successive-Cancellation for the Slow Transform}
\label{sec:scslow}
The original decoding algorithm of polar codes is successive-cancellation. 
More generally, SC is also employed for encoding~\cite{Honda_Yamamoto_2013}. 
Better coding results can be obtained using SCL decoding. 
However, we first discuss SC decoding. 

The universal scheme~\cite{sasoglu_2016_universal,Shuval_Tal_Universal} employs a joint transform consisting of slow and fast transforms. The joint transform is recursive as well, and hence conveniently described via hyperlayers, hyperbranches, and hyperphases --- to be detailed in the full paper.
SC can be used for encoding and decoding this joint transform. 
Due to length constraints, we focus our discussion on SC for the slow transform --- a cardinal building block. 
We remark that the slow transform, used exclusively, is not sufficient for coding as it polarizes too slowly. 
Our full paper will provide details on the joint transform. 

SC is used in a probabilistic setting. 
Denote random variables using capital letters. 
For channel coding, $X_0^{N-1}$ is the channel input and $Y_0^{N-1}$ is the corresponding output. 
Their joint probability is governed by a hidden Markov state chain (see~\cite{Shuval_Tal_Universal} for full details of the model):
$\Probi{X_i, Y_i, S_i | S_{i-1}}$, where $S_i$ is the state at time $i$. 
The states belong to a finite set $\mathcal{S}$. We also denote $\mathcal{X} = \{0,1\}$.

Algorithm~\ref{alg:highLevel} is a general high-level description of SC for the above setting. For each of the $N$ phases, we first compute 
\begin{IEEEeqnarray}{l}
         p(u, s, s'; \hat{u}_0^{\varphi-1}, y_0^{N-1}) \IEEEyesnumber \label{eq_def of p(u,s,s';uhat,y)} \\
        \!\triangleq\! \Prob{\ns U_\varphi = u, S_{-1} =  s, S_{N-1} =  s';\,  U_0^{\varphi-1} =  \hat{u}_0^{\varphi-1}, Y_0^{N-1}= y_0^{N-1}\ns}\ns , \IEEEnonumber
\end{IEEEeqnarray}
where $U_0^{N-1}$ is the transform of $X_0^{N-1}$. 
Every phase $\varphi$ is either used to carry information bits (such a phase is called a \emph{data phase}) or not. 
In~\cite{Arikan_2009}, non-data phases were called `frozen.' More generally~\cite{Honda_Yamamoto_2013}, these are shaping phases. 
Either way, for non-data phases $u_{\varphi}$ is determined via a mapping\footnote{Note that the Cyclic Redundancy Check (CRC) variant~\cite{Tal_2015} of polar codes places CRC bits in certain phases. Under our definition, these are also shaping phases. The same comment applies for the dynamically frozen bits of~\cite{Trifonov_CRC}. }  $\mathcal{F}_{\varphi}(u_0^{\varphi-1})$.
In~\cite{Honda_Yamamoto_2013}, this mapping utilizes common randomness between encoder and decoder. 
When $\varphi$ is a data phase, we determine $\hat{u}_{\varphi}$ using a maximum aposteriori criterion. 
That is, we compute 
\begin{equation} 
    p(u; \hat{u}_0^{\varphi-1}, y_0^{N-1}) \triangleq \sol{\sum_{s, s' \in \mathcal{S}}} p(u, s, s' ; \hat{u}_0^{\varphi-1}, y_0^{N-1}), 
    \label{eq_def of p(u,s,s')}
\end{equation} 
and select $\hat{u}_{\varphi} = \arg\max_{u\in\mathcal{X}} p(u; \hat{u}_0^{\varphi-1}, y_0^{N-1})$.

\begin{algorithm}
\SetInd{0.49em}{0.49em}
\caption{A high-level description of SC}
\label{alg:highLevel}
\KwIn{received $y_0^{N-1}$ (or empty vector for encoding)}
\KwOut{transformed codeword $\hat{u}_0^{N-1}$}
\BlankLine
\For{$\varphi = 0,1,\ldots, N-1$}
{
    compute $p(u, s, s'; \hat{u}_0^{\varphi-1}, y_0^{N-1})$ for $u \in \mathcal{X}$; $s, s' \in  \mathcal{S}$ \;
    \eIf{$\varphi$ is a frozen (shaping) phase}
  {
      set $\hat{u}_{\varphi} \gets \mathcal{F}_{\varphi}(\hat{u}_0^{\varphi-1})$ \; 
  }
  {
      set $\hat{u}_{\varphi} \gets \arg\max_{u\in \mathcal{X}} p( u ; \hat{u}_0^{\varphi -1}, y_0^{N-1})$
  }
}
\Return $\hat{u}_0^{N-1}$
\end{algorithm}

\subsection{A first implementation of Algorithm~\ref{alg:highLevel}}
Our first implementation mirrors Algorithms 1 -- 4 of~\cite{Tal_2015}, modified and generalized to the slow transform and to settings with memory. 
Thus, we first employ straightforward data structures: arrays. 
Later, when considering list decoding, we will show that the space complexity can be reduced using enhanced data structures. 

To implement Algorithm\nobreakspace \ref {alg:highLevel}, we need a way to compute $p(u,s, s' ; \hat{u}_0^{\varphi-1}, y_0^{N-1})$. 
This is accomplished using the recursive description~\eqref{eq_Fj for lateral indices_pbl} and~\eqref{eq_Fj for medial indices_pbl}.
The intermediate calculations required for computing $p(u, s, s'; \hat{u}_0^{\varphi-1}, y_0^{N-1})$ are common between different phases $\varphi$. 
We store some of the intermediate calculations for time complexity reduction. 

Our implementation utilizes two main data structures: one for keeping track of intermediate bit decisions and the other for storing intermediate probabilities. Specifically, for each layer $0 \leq \lambda \leq n$ we define a \emph{bit-decision array} $B_\lambda$ of size $N = 2^\lambda N_0 \cdot 2^{n-\lambda}$. The array starts out uninitialized, and when the algorithm concludes it holds bit decisions:
\[
    B_\lambda[\pb{\varphi}{\beta}] = \hat{u}^{(\lambda)}_{\pb{\varphi}{\beta}}.
\]
We further define, for each layer $\lambda$, a \emph{probabilities array} $P_\lambda$ of size $N \times |\mathcal{X}| \times |\mathcal{S}| \times |\mathcal{S}|$. When the algorithm concludes,  
\begin{equation*} \label{eq_Plambda def}
    P_{\lambda}[\pb{\varphi}{\beta}][u,s,s'] = \plam{\lambda}{\varphi}{\beta}(u,s,s'),
\end{equation*}
where we define $\plam{\lambda}{\varphi}{\beta}(u,s,s')$ in~\eqref{eq_def of plam}. For brevity, we denote $\Lambda = 2^{\lambda}N_0$,
$\tilde{S} =  S_{\beta\Lambda -1}$, $\tilde{S}' =  S_{(\beta+1)\Lambda -1}$,   $v_{\varphi} = u^{(\lambda)}_{\pb{\varphi}{\beta}}$, $\hat{v}_{\varphi}= \hat{u}^{(\lambda)}_{\pb{\varphi}{\beta}}$,  and $\tilde{y}_{\varphi} = y_{\pbl{\varphi}{\beta}{\lambda}}$. Capital versions of $v_{\varphi}$ and $\tilde{y}_{\varphi}$ denote random variables. Then, 
\begin{IEEEeqnarray}{l}
    \plam{\lambda}{\varphi}{\beta}(u,s,s') \IEEEyesnumber \label{eq_def of plam} \\ 
    \quad = \Prob{V_{\varphi} = u, \tilde{S} = s, \tilde{S}' = s' ; V_0^{\varphi-1} =\hat{v}_0^{\varphi-1}, \tilde{Y}_0^{\Lambda-1} = \tilde{y}_0^{\Lambda- 1} }. \IEEEnonumber
\end{IEEEeqnarray} 
Observe that when $\lambda = 0$ then $V_{\varphi} = X_{\varphi}$, the channel input, and $\Lambda = N_0$. 
Thus, $\plam{0}{\varphi}{\beta}$ involves a sub-vector of the output $Y_0^{N-1}$ of size $N_0$. 
Due to the Markov property, we can compute $\plam{0}{\varphi}{\beta}$ directly from the joint distribution $\Probi{X_i,Y_i,S_i | S_{i-1}}$. 
Observe that when $\lambda = n$, $ \plam{n}{\varphi}{\beta}(u,s,s') = p(u,s,s';\hat{u}_0^{\varphi-1},y_0^{N-1})$.

Our implementation is given in Algorithm\nobreakspace \ref {alg:firstImplementation_main}. 
Its main loop (lines~\ref{alg:mainloop} -- \ref{alg:setUprop2}) iterates over all phases of the single branch of the last layer $n$. 
For each last-layer phase it recursively calculates relevant probabilities of the probabilities
array (line~\ref{alg:recursivelyCalcP}), decides on the value of the last-layer
phase (lines~\ref{alg:decideU1} -- \ref{alg:decideU2}), and finally propagates this value throughout
the transform (lines~\ref{alg:setUprop1} -- \ref{alg:setUprop2}). 

An additional array in our implementation is the \emph{tracker} $T_{\lambda}[i]$, $i \in\{0,1\}$.
For each layer $0 \leq \lambda< n$, each of its two elements is either empty or holds a phase-branch pair
$(\bar{\varphi}, \bar{\beta})$. A \resetTracker function sets all of its elements over all layers to empty. 
Whenever $B_{\lambda}[\pb{\varphi}{\beta}]$ is updated, the tracker is also updated. 
As will soon become apparent, no more than two phase-branch pairs $(\bar{\varphi}, \bar{\beta})$ are updated per layer $\lambda$ in an iteration of the
main loop of Algorithm\nobreakspace \ref {alg:firstImplementation_main}. 
Thus, $T_{\lambda}[i] = (\bar{\varphi}, \bar{\beta})$ means that in the previous iteration of the main
loop, phase $\bar{\varphi}$ of branch $\bar{\beta}$ in layer $\lambda$ was updated. 

We will soon see that recursive probability calculation  needs to know which phases and branches were updated in every layer in the previous iteration of the main loop of~Algorithm\nobreakspace \ref {alg:firstImplementation_main}. 
This is accomplished via the tracker array. 
Specifically, before propagating bit decisions throughout the transform, we reset the tracker (line~\ref{alg:tracker reset} in Algorithm\nobreakspace \ref {alg:firstImplementation_main}).

\begin{algorithm}
\caption{First implementation of SC decoder}
\label{alg:firstImplementation_main}
\KwIn{received $y_0^{N-1}$ (empty vector for encoding)}
\KwOut{transformed codeword $\hat{u}_0^{N-1}$}
\BlankLine
\For(\tcp*[h]{Initialization}){$\beta = 0,1,\ldots, 2^n-1$}
{
        \For{$u \in \mathcal{X}, s \in \mathcal{S}, s' \in \mathcal{S}$}
        {
            set $\varphi \gets 0 $ 
            \tcp*[h]{Other phases updated later}\; 
            $P_{0}[\pb{\varphi}{\beta}][u,s,s'] \gets \plam{0}{\varphi}{\beta}(u,s,s')$  \label{alg_p0 initialization}\; 
}
}
$\resetTracker()$\;
set $\beta \gets 0$ \tcp*[h]{The only branch of layer $n$}\; 
\For(\tcp*[h]{Main loop}){$\varphi = 0,1,\ldots, N-1$ \label{alg:mainloop}}
{ 
  $\recursivelyCalcP(n, \varphi, \beta)$\; \label{alg:recursivelyCalcP}
  \eIf{$\varphi$ is a frozen (shaping) phase \label{alg:decideU1}}
  {
      set $\hat{u}_{\varphi} \gets \mathcal{F}_{\varphi}(\hat{u}_0^{\varphi-1})$ \; 
  }
  {
      set $\hat{u}_{\varphi} \gets \arg\max_{u\in \mathcal{X}} p( u ; \hat{u}_0^{\varphi -1},
      y_0^{N-1})$ \label{alg:decideU2}
}
set $B_n[\pb{\varphi}{\beta}] \gets \hat{u}_{\varphi}$ \label{alg:setUprop1} \; 
$\resetTracker()$ \label{alg:tracker reset} \; 
$\recursivelyUpdateB(n, \varphi, \beta)$\label{alg:setUprop2}\;
}
\Return $\hat{u}_0^{N-1}$
\end{algorithm}

Algorithm\nobreakspace \ref {alg:firstImplementation_calcP}, invoked with $\recursivelyCalcP(\lambda,\varphi,\beta)$, computes $\plam{\lambda}{\varphi}{\beta}(u,s,s')$ for all $u \in \mathcal{X}$ and $s,s' \in \mathcal{S}$. 
It does this by utilizing the relationships in equations~\eqref{eq_Fj for lateral indices_pbl} and~\eqref{eq_Fj for medial indices_pbl} and the Markov property. 
However, it must first ensure that the relevant indices in $P_{\lambda-1}$ had been computed. 
The branches and phases in  layer $\lambda-1$ for which $P_{\lambda-1}$ is updated depend on $\lambda$, $\varphi$, and $\beta$. 
There are three cases. 
\begin{enumerate}
    \item When $\lambda = 0$ (line~\ref{alg:lam0}),  we update the base probabilities if needed. 
Efficient implementation, especially when list decoding is involved, requires another algorithm, \updateBaseProbs, see our full paper. 
\item  If $\lambda>0$ and $\varphi = 0$ (line~\ref{alg:varphi0}), no bit decisions had been propagated, and we recurse to the previous layer.  
\item Otherwise, we use the tracker array $T_{\lambda-1}$ (line~\ref{alg:varphi+1}). 
Note that this calls upon a `next phase' $\bar{\varphi}+1$ in a branch. 
This `next phase'  may exceed the size of the branch. 
This happens when all phases in a branch had their bit decisions propagated to. 
Thus (lines~\ref{alg:phiexceedsInit}--\ref{alg:phiexceedsFin}) we set to zero $\plam{\lambda-1}{\bar{\varphi}}{\bar{\beta}}(u,\sigma,\sigma')$ for all $u \neq B_{\lambda-1}[\pb{\bar{\varphi}}{\bar{\beta}}]$, and $\sigma,\sigma' \in \mathcal{S}$. 
\end{enumerate}

Finally, in lines~\ref{alg:computeProbsInit}--\ref{alg:computeProbsFin}, $\recursivelyCalcP(\lambda,\varphi,\beta)$ computes $\plam{\lambda}{\varphi}{\beta}(\cdot,\cdot,\cdot)$.
The computation depends on the type of phase $\varphi$: lateral or medial, and uses equations~\eqref{eq_Fj for lateral indices_pbl} and~\eqref{eq_Fj for medial indices_pbl}, respectively. 
It also relies on the Markov property and the adaptation of minus and plus transforms to this case, see~\cite{wang2014joint,Wang_2015}. 
Details of \lateralProbHelper and \medialProbHelper will appear in our full paper. 

\begin{algorithm} 
\SetInd{0.52em}{0.52em}
\caption{$\protect\recursivelyCalcP(\lambda, \varphi, \beta)$}
\label{alg:firstImplementation_calcP}
\KwIn{$\lambda$ = layer, $\beta =$ branch in layer, $\varphi =$ phase in branch}
\BlankLine
\If(\tcp*[h]{after last phase in branch}){$\varphi =  2^{\lambda}N_0$ \label{alg:phiexceedsInit}}
{
    \For{$u \in \mathcal{X}$} 
    {
    \tcp*[h]{Set probability zero to u different than last decision}\;
    \If{ $u \neq B_{\lambda}[\pb{\varphi -1}{\beta}]$}
        {
    \lFor{$s,s' \in\mathcal{S}$}
    { 
        $P_{\lambda}[\pb{\varphi-1}{\beta}][u,s,s'] \gets 0$  \label{alg:phiexceedsFin}
    }
}
}
    \Return\; 
}
\If{$\lambda = 0$  \label{alg:lam0}}
{
    \lIf{$\varphi > 0$}
    {
    $\updateBaseProbs(\varphi, \beta)$ 
    }
\Return\tcp*[h]{Stopping condition}\;
}
\If{$\varphi = 0$ \label{alg:varphi0}}
{
     $\recursivelyCalcP(\lambda-1, \varphi, 2\beta)$  \; 
     $\recursivelyCalcP(\lambda-1, \varphi, 2\beta+1)$   
}
\Else(\tcp*[h]{$\varphi > 0$})
{
    \For{$i \in \{0,1\}$} 
    {\If{$T_{\lambda-1}[i]$ is not empty}
        {
            $(\bar{\varphi}, \bar{\beta}) \gets T_{\lambda-1}[i]$\;  
            \tcp*[h]{Prepare next phase in branch}\; 
            $\recursivelyCalcP(\lambda-1, \bar{\varphi}+1, \bar{\beta})$ \label{alg:varphi+1} \; 
        }
    }
}
\tcp*[h]{Compute $P_{\lambda}[\pb{\varphi}{\beta}][u,s,s']$ for all $u,s,s'$}\; 
\If{$\varphi \in \latset{\lambda}$ \label{alg:computeProbsInit}}
{
    $\lateralProbHelper(\lambda,\varphi,\beta)$ \tcp*[h]{use \eqref{eq_Fj for lateral indices_pbl}}                \; 
}
\ElseIf{$\varphi \in \medset{\lambda}$}
{
    $\medialProbHelper(\lambda,\varphi,\beta)$  \tcp*[h]{use \eqref{eq_Fj for medial indices_pbl}}\label{alg:computeProbsFin} \;
}
\Return
\end{algorithm}

Algorithm\nobreakspace \ref {alg:firstImplementation_updateB} resolves equations~\eqref{eq_Fj for lateral indices_pbl} and~\eqref{eq_Fj for medial indices_pbl} recursively: 
it is invoked after $B_{\lambda}[\pb{\varphi}{\beta}]$ had been set, and propagates this throughout the transform.
The recursive computation is performed from layer $\lambda$ to layer $\lambda-1$, the opposite
direction to that of equations~\eqref{eq_Fj for lateral indices_pbl} and~\eqref{eq_Fj for medial indices_pbl}. 
When $\lambda=0$, $\recursivelyUpdateB(\lambda,\varphi,\beta)$ cannot propagate to a previous layer, so it simply returns. 
Otherwise, its operation depends on the type of phase $\varphi$. 
\begin{enumerate}
    \item $\varphi \in \latset{\lambda}$ (lines~\ref{alg:calcBlatinit}--\ref{alg:calcBlatfin}). By~\eqref{eq_Fj for lateral indices_pbl}, a  lateral phase of layer $\lambda$ passes-through directly to a single phase of layer $\lambda-1$.  
        Thus, only a single phase-branch pair of $B_{\lambda-1}$ is updated, and the algorithm recurses to layer $\lambda-1$. After every update of the bit-decision array  we also update the tracker. In this case, $T_{\lambda-1}$ has only one non-empty entry. 
    \item $\varphi \in \medsetMinus{\lambda}$ (line~\ref{alg:calcBmedMinus}). Medial phases come in minus and plus pairs, in this order. Both members of the pair are required to resolve~\eqref{eq_Fj for medial indices_pbl} for $\lambda-1$. Since this is the first member of the pair, we must wait. Nothing is updated, so return without recursing.
    \item $\varphi \in \medsetPlus{\lambda}$
        (lines~\ref{alg:calcBmedPlusinit}--\ref{alg:calcBmedPlusFin}). We now have the left-hand
        side of~\eqref{eq_Fj for medial indices_pbl} for two consecutive phases, a minus and a plus
        pair (for $\varphi-1$ odd and $\varphi$ even), and can resolve for the right-hand side, namely layer $\lambda-1$. Two phase-branch pairs of $B_{\lambda-1}$ are updated, and entered into the tracker $T_{\lambda-1}$. The algorithm recurses for these two pairs. 
\end{enumerate}

\begin{algorithm}
\caption{$\protect\recursivelyUpdateB(\lambda,\varphi,\beta)$}
\label{alg:firstImplementation_updateB}
\KwIn{$\lambda$ = layer, $\beta =$ branch in layer, $\varphi =$ phase in branch}
\BlankLine
\If{$\lambda = 0$ or $\varphi \in\medsetMinus{\lambda}$ \label{alg:calcBmedMinus}}
{
    \Return \tcp*[h]{Only medial plus or lateral phases are propagated} \; 
}
\If{$\varphi \in \latset{\lambda}$ \label{alg:calcBlatinit}}
{
set $\psi \gets \lfloor\varphi/2 \rfloor$\; 
set $\bar{\beta} \gets 2\beta + (\varphi\mod 2)$ \tcp*[h]{See \eqref{eq_Fj for lateral indices_pbl}}\; 
$B_{\lambda-1}[\pb{\psi}{\bar{\beta}}] \gets B_{\lambda}[\pb{\varphi}{\beta}]$\; 
$T_{\lambda-1}[0] \gets (\psi,\bar{\beta})$ \;
$\recursivelyUpdateB(\lambda-1, \psi, \bar{\beta})$ \label{alg:calcBlatfin} \; 
}
\Else(\tcp*[h]{ See \eqref{eq_Fj for medial indices_pbl}} \label{alg:calcBmedPlusinit})
{ 
    set $\psi' \gets \lfloor (\varphi - 1)/2 \rfloor$\; 
    \If{$\varphi \in \medsetPlus{\lambda}$}
{
    \If{$\psi' \in \medsetMinus{\lambda-1}$}
    {
        $B_{\lambda-1}[\pb{\psi'+1}{2\beta}] \gets
        B_{\lambda}[\pb{\varphi}{\beta}]+B_{\lambda}[\pb{\varphi-1}{\beta}]$ \label{alg:changethisline1}\; 
        $B_{\lambda-1}[\pb{\psi'}{2\beta+1}] \gets B_{\lambda}[\pb{\varphi}{\beta}]$\; 
    }
    \Else
    {
        $B_{\lambda-1}[\pb{\psi'}{2\beta+1}] \gets
        B_{\lambda}[\pb{\varphi}{\beta}]+B_{\lambda}[\pb{\varphi-1}{\beta}]$ \label{alg:changethisline2}\; 
        $B_{\lambda-1}[\pb{\psi'+1}{2\beta}] \gets B_{\lambda}[\pb{\varphi}{\beta}]$\; 
    }
$T_{\lambda-1}[0] \gets (\psi',2\beta+1)$ \;
$T_{\lambda-1}[1] \gets (\psi'+1,2\beta)$ \;
\BlankLine
$\recursivelyUpdateB(\lambda-1, \psi', 2\beta+1)$  \label{alg:changethisline3}\; 
$\recursivelyUpdateB(\lambda-1, \psi'+1, 2\beta)$ \label{alg:calcBmedPlusFin} \label{alg:changethisline4}\; 
    
}
}
\end{algorithm}

Algorithm\nobreakspace \ref {alg:firstImplementation_updateB} highlights an important observation on the slow transform. 
Recall from~\eqref{eq_Ln}, \eqref{eq_defs of lat-med sets}, and \eqref{eq_Fj for medial indices_pbl} that consecutive medial phases of layer $\lambda$ are formed from two medial phases of layer $\lambda-1$. 
One of these phases is in $\medsetMinus{\lambda-1}$ and the other is in $\medsetPlus{\lambda-1}$. 
Thus, when the algorithm is invoked for phase $\varphi \in \medsetPlus{\lambda}$ it recursively invokes the algorithm twice, once for a phase in $\medsetMinus{\lambda-1}$ and once for a phase in $\medsetPlus{\lambda-1}$. 
Hence, when Algorithm\nobreakspace \ref {alg:firstImplementation_main} invokes $\recursivelyUpdateB(n,\varphi,0)$ for $\varphi \in \medsetPlus{n}$, two entries of the bit-decision array will be updated for \emph{every} layer $0 \leq \lambda \leq n-1$. 
This is in stark contrast to the fast transform (which typically updates the bit-decision array for only a few layers), and the reason that the list decoder implementation of~\cite{Tal_2015} does not carry through. 

The following lemma will be crucial for the list decoder's bookkeeping. 
To this end, we first define a \emph{branc}.
\begin{definition}[branc] 
    A \emph{branc} contains two consecutive branches. The branc of branch $\beta$ is numbered $\lfloor \beta/2 \rfloor$. 
    In other words, branches $\beta$ and $\beta+1$ are in the same branc if their bit-expansions are
    equal up to the least significant bit.\footnote{A `branc' is a `branch' whose ``least
    significant letter,'' `h',  is dropped.}
\end{definition}
For example, there are eight brancs for layer $\lambda=4$: branc $0=\langle000\rangle_2$ contains branches $0= \langle0000\rangle_2$ and $1=\langle0001\rangle_2$, branc $1$ contains branches $2$ and $3$, etc. 

We order brancs in bit-reversed cyclic order. Thus, for $\lambda=4$, brancs are ordered: $0=\langle000\rangle_2$, $4=\langle100\rangle_2$, $2=\langle010\rangle_2$, $6=\langle110\rangle_2$, $1=\langle001\rangle_2$, $5=\langle101\rangle_2$, $3=\langle011\rangle_2$, $7=\langle111\rangle_2$. Since the order is cyclic, the next branc after $7$ is $0$. 

We say that a branc of $B_{\lambda}$ is updated if $B_{\lambda}$ is updated for at least one of the branches $\beta$ in the branc. 
Namely, $B_{\lambda}[{\varphi, \beta}]$ is updated for some phase $\varphi$ and branch $\beta$ in the branc. 
A similar definition holds for $P_{\lambda}$. 
\begin{lemma}\label{lem_cyclic B}
    For each layer $\lambda$, the brancs of $B_{\lambda}$ are updated  in bit-reversed cyclic order during the entire run of
    Algorithm\nobreakspace \ref {alg:firstImplementation_main}.
\end{lemma}
\begin{corollary}\label{cor_cyclic P}
    For each layer $\lambda$, the brancs of $P_{\lambda}$ are updated  in bit-reversed cyclic order during the entire run of the main loop of 
    Algorithm\nobreakspace \ref {alg:firstImplementation_main}, save for the first iteration, $\varphi = 0$. 
\end{corollary}

The following theorem reduces the space complexity. 
\begin{theorem}\label{thm_no phase for BP}
    Algorithms\nobreakspace  \ref {alg:firstImplementation_main} to\nobreakspace  \ref {alg:firstImplementation_updateB} 
    can be implemented with per-layer bit-decision arrays and probabilities arrays indexed only by
    branch. 
\end{theorem}
I.e., the bit-decision array can be indexed as $B_{\lambda}[\beta]$ and
the probabilities array as $P_{\lambda}[\beta](u,s,s')$. 
Their entries will refer to the last updated phase in the relevant layer and branch. 
Note that this entails changing the interface of \recursivelyUpdateB to also pass the bit-decision of the previous phase. 
Namely, lines~\ref{alg:changethisline1}, \ref{alg:changethisline2}, \ref{alg:changethisline3}, \ref{alg:changethisline4} of Algorithm\nobreakspace \ref {alg:firstImplementation_updateB} must be changed. 

\section{List Decoding and the Cyclic Exponential Array}
List decoding  uses a number, $\mathcal{L}$, of decoding paths. 
A path up to phase $\varphi$ is split at the decision point (the bit decision for phase $\varphi$ in the last 
layer $n$, when $\varphi$ is a data phase). 
Thus, the number of paths is doubled at every split. 
If this number exceeds $\mathcal{L}$, we prune the list and keep the $\mathcal{L}$ most likely paths.

The paths differ in their bit decisions in several places, and consequently the arrays $B_{\lambda}$ and $P_{\lambda}$ differ for different paths. 
The essence of an \emph{efficient} implementation of list decoding is to share portions of these arrays among paths. 
Namely, if a portion of an array is the same for two paths, we store it in memory once. 
Conversely, array portions that are to be written to by a path at the current phase must not be shared. 

The universal transform is formed by concatenating a sequence of slow transforms and a final fast
transform. 
A key building block in list decoding for the universal transform is the \emph{cyclic exponential array} (CEA). 
This is a data structure that enables sharing array portions efficiently. 
We focus on the slow transform, due to space limitations. 

Recall from Theorem\nobreakspace \ref {thm_no phase for BP} that the arrays $B_{\lambda}$ and $P_{\lambda}$ may be indexed by branch only. 
Further note from Lemma\nobreakspace \ref {lem_cyclic B} and Corollary\nobreakspace \ref {cor_cyclic P} that these arrays are updated in a cyclic order. 

The CEA data structure holds generic objects:
\begin{itemize}
    \item For $B_{\lambda}$ it holds pairs of bit decisions, one pair for each branc. 
    \item For $P_{\lambda}$ it holds pairs of `probability datums,' one pair for each branc. 
A probability datum holds $|\mathcal{X}|\cdot|\mathcal{S}|^2$ probabilities, indexed by $u$, $s$, and $s'$.
\end{itemize}
A CEA contains $2^{\lambda}$ objects for some $\lambda$. 
The CEA supports two operations: $\algread(i)$ and $\algwrite(i)$. 
The operation $\algread(i)$ returns the object stored at position $i$ of the CEA. 
The operation $\algwrite(i)$ stores an object at position $i$ of the CEA. 
The first \algwrite must be called with index $i=0$. 
For subsequent calls, if the previous call of \algwrite was with index $i$, the current call must be
with either index $i$ or $i+1$, modulo the CEA size $2^{\lambda}$.
Bit-reversal is performed by the caller. 

Internally, a CEA of size $2^{\lambda}$ holds the following variables:
\begin{itemize}
    \item \texttt{lastIndexWrittenTo}: the last index written to by \algwrite.  
    \item \texttt{lastWrittenValue}: the object last written by \algwrite. 
    \item Arrays $\texttt{currentCycleArray}_\tau$, $0 \leq \tau < \lambda$. Array $\texttt{currentCycleArray}_{\tau}$ holds $2^{\tau}$ objects; its indexing is zero-based. 
    \item \texttt{previousCycleArray}: holds $2^{\lambda}$ objects; its indexing is zero-based. 
\end{itemize}
When $\texttt{lastIndexWrittenTo} = i$, 
\begin{itemize}
    \item \texttt{lastWrittenValue} holds the object written to by the latest write, $\algwrite(i)$. 
    \item For $j > i$, $\texttt{previousCycleArray}[j]$ holds the object written to by the latest $\algwrite(j)$. 
    \item For $j < i$, the object written to by the latest $\algwrite(j)$ is in $\texttt{currentCycleArray}_{\tau}[k]$, where $\tau$ and $k$ are computed as follows. 
        Let $i = \langle i_{\lambda-1} i_{\lambda-2}\cdots i_{0}\rangle_2$  and $j = \langle j_{\lambda-1} j_{\lambda-2}\cdots j_{0}\rangle_2$ be the binary bit expansions of $i$ and $j$ respectively, with $i_0$ ($j_0$) the least significant bit of $i$ ($j$). Then, $\tau$ is the largest integer such that $i_{\tau} = 1$ and $j_{\tau} = 0$, and $k = \langle j_{\tau}j_{\tau-1} \cdots j_0\rangle_2$. 
\end{itemize}                                                                                 

\begin{example}
    Let $\lambda = 4$ and $\texttt{lastIndexWrittenTo} = 11 = \langle 1011 \rangle_2$. 
    Then, for $0 \leq j \leq 15$, the location that $\algread(j)$ will access is the cell numbered $j$ in the following:
    \begin{align*}
        \texttt{currentCycleArray}_3 &\equiv \begin{bmatrix} 0 & 1 & 2 & 3 & 4 & 5 & 6 & 7  \end{bmatrix}, \\
        \texttt{currentCycleArray}_2 &\equiv \begin{bmatrix} \text{N/A} & \text{N/A} & \text{N/A} & \text{N/A}\end{bmatrix}, \\
        \texttt{currentCycleArray}_1 &\equiv \begin{bmatrix} 8 & 9 \end{bmatrix}, \\
        \texttt{currentCycleArray}_0 &\equiv \begin{bmatrix} 10 \end{bmatrix}, \\
        \texttt{lastWrittenValue} &\equiv  11, \\
        \texttt{previousCycleArray} &\equiv \begin{bmatrix} \text{N/A}\!\!\ns&\!\ns \cdots \!\ns &\!\! \ns\text{N/A}\! &\! 12 &\! 13 &\! 14 &\! 15 \end{bmatrix}\!\ns.
    \end{align*}
    The only legal \algwrite operations are $\algwrite(11)$ and $\algwrite(12)$: $\algwrite(11)$ changes \texttt{lastWrittenValue} only; 
    $\algwrite(12)$ first copies into $\texttt{currentCycleArray}_2$ objects from \texttt{lastWrittenValue} and  $\texttt{currentCycleArray}_{\tau}$, $\tau=0,1$.
    Then, it changes \texttt{lastWrittenValue} and \texttt{lastIndexWrittenTo}.
    Crucially, the two largest arrays, \texttt{previousCycleArray} and $\texttt{currentCycleArray}_3$ are unchanged. 
\end{example}

The following theorem details which internal variables of a CEA are changed during a \algwrite operation. 
This is crucial with respect to list decoding. 
Namely, it details which variables can be shared among paths after a split, and which cannot. 
To this end, the binary bit expansions of $i$ and $j$ respectively are $i = \langle i_{\lambda-1} i_{\lambda-2}\cdots i_{0}\rangle_2$  and $j = \langle j_{\lambda-1} j_{\lambda-2}\cdots j_{0}\rangle_2$. 

\begin{theorem} \label{thm:CEAwrite}
    Let $\texttt{lastIndexWrittenTo} = i$ and consider $\algwrite(j)$ for $j=(i+1) \mod 2^{\lambda}$. 
    Apart from \texttt{lastWrittenValue} and \texttt{lastIndexWrittenTo}, a single array is changed: 
    \begin{itemize}
        \item If $j=0$, only \texttt{previousCycleArray} is changed. 
        \item Otherwise, let $\tau$ be the largest integer such that $j_{\tau} = 1$ and $i_{\tau} = 0$. 
            Then, only $\texttt{currentCycleArray}_{\tau}$ is changed. 
    \end{itemize}
\end{theorem}

\begin{corollary}
    Let $i$, $j$, and $\tau$ be as in Theorem\nobreakspace \ref {thm:CEAwrite}, and let $2^{\lambda}$ be the CEA size. Then the time complexity of $\algwrite{j}$ is $O(2^{\lambda})$ if $j = 0$ and $O(2^{\tau})$ otherwise. 
    Thus, a sequence of $2^{\lambda}$ write operations spanning all indices $j$ takes time $O(\lambda\cdot 2^{\lambda})$. 
\end{corollary}
\vspace{-0.1cm}

The number of layers $n+1$ in a slow transform of blocklength $N = 2^{n}N_0$ is $O(\log N)$. Thus, \vspace{-0.1cm} 
\begin{corollary}\label{cor:slowComplexity}
    SCL for a slow transform of length $N$ can be accomplished with space complexity $O(\listSize\cdot N)$ and time complexity $O(\listSize\cdot N \log^2N)$.
\end{corollary}
\vspace{-0.1cm}

One might infer from Corollary\nobreakspace \ref {cor:slowComplexity} that the overall time complexity of SCL decoding of a universal transform of blocklength $N$ is $O(\mathcal{L}\cdot N \log^2N)$.
This would happen in a straightforward implementation. 
However, the universal transform has a parallel structure, in which multiple identical slow transforms are decoded in lockstep. 
This allows for significant savings in time complexity, by having the CEA objects be pointers to arrays whose length is the number of slow transform copies. 
Copying an object is simply copying a pointer, and a single copy operation suffices for all parallel slow transforms. 
Hence, the bookkeeping associated with all parallel slow transforms is not a function of the number of parallel transforms, only the blocklength of a single slow transform. 
Thus, it is possible to show the following. 

\begin{theorem}
    SCL for a universal transform of blocklength $N$  can be accomplished with space complexity $O(\listSize\cdot N)$ and time complexity $O(\listSize\cdot N \log N)$.
\end{theorem}

\twobibs{
\bstctlcite{BIBctrl}
\bibliographystyle{IEEEtran} 
\bibliography{../mybib.bib} 
}
{

}

\end{document}